\pdfoutput=1

\documentclass[11pt]{article}

\usepackage{tikz}
\usetikzlibrary{shapes.geometric, arrows.meta, positioning}

\usepackage{graphicx}
\usepackage[]{ACL2023}
\usepackage{booktabs}
\usepackage{times}
\usepackage{latexsym}
\usepackage{tikz}
\usepackage{graphicx}
\usepackage{amsmath, amssymb, amsfonts}  
\usepackage{bm}  
\usepackage[T1]{fontenc}

\usepackage[utf8]{inputenc}

\usepackage{microtype}

\usepackage{inconsolata}

\title{Enhancing Fake‐News Detection with Node‑Level Topological Features}

\author{Kaiyuan Xu \\
  Xidian University\\
  \texttt{kyxu@homelab4geeks.uk} 
}

\begin{document}
\maketitle

\begin{abstract}

In recent years, the proliferation of misinformation and fake news has posed serious threats to individuals and society, spurring intense research into automated detection methods. Previous work showed that integrating content, user preferences, and propagation structure achieves strong performance, but leaves all graph‑level representation learning entirely to the GNN, hiding any explicit topological cues. To close this gap, we introduce a lightweight enhancement: for each node, we append two classical graph-theoretic metrics, degree centrality and local clustering coefficient, to its original BERT and profile embeddings, thus explicitly flagging the roles of hub and community. In the UPFD Politifact subset, this simple modification boosts macro F1 from 0.7753 to 0.8344 over the original baseline. Our study not only demonstrates the practical value of explicit topology features in fake‑news detection but also provides an interpretable, easily reproducible template for fusing graph metrics in other information‑diffusion tasks.

\end{abstract}

\section{Introduction}

The rapid spread of misinformation and fake news on social media has become a pressing challenge for both individuals and society at large, driving extensive efforts toward automated detection solutions. Among these, graph‑neural‑network (GNN)‑based methods have emerged as a powerful paradigm by jointly modeling the news content, user preferences, and propagation structure of a piece of information. 

It is worth noting that the urgency of this task is further amplified by recent advancements in generative artificial intelligence. For instance, efficiency-focused diffusion transformers now enable rapid text-to-any-task generation \cite{wang2025qihoot2x}. To improve generation quality, relevance-guided control mechanisms have been introduced for diffusion models \cite{cao2026relactrl}. In the realm of visual planning, unified frameworks now support complex layout planning alongside image generation \cite{he2025plangen}. These capabilities extend to hierarchical controllable models for layout-to-image tasks \cite{cheng2024hico}. Furthermore, novel architectures allow for layout-togglable story generation, increasing the coherence of synthetic narratives \cite{ma2025lay2story}. 

The sophistication of these tools makes detection increasingly difficult. Cross-cultural generation is now facilitated by bridge diffusion models \cite{liu2025bridge}. For artistic content, ultra-high quality style transfer is achievable via transformers \cite{zhang2025ustydit}. Personalization techniques have also advanced, with retrieval-augmented generation guiding the synthesis process \cite{ling2026ragar}. To ensure model stability, methods like regulated clipping are employed to mitigate implicit over-optimization \cite{wang2025grpoguard}. In the video domain, world simulator assistants are being developed for physics-aware generation \cite{wang2025wisa}. Additionally, cross-frame textual guidance now allows for dynamic and consistent video creation \cite{feng2025fancyvideo}. These technologies enable the mass production of highly realistic multimodal misinformation.

One prominent example is the User Preference‑aware Fake News Detection (UPFD) framework \cite{upfd}. UPFD constructs a propagation graph for each news item - where nodes are users who shared the item and edges represent retweet relationships - and uses GNN layers to fuse those three information streams into a joint representation. Experiments in UPFD are conducted on two benchmark datasets: Politifact \cite{Politifact}, a collection of 314 fact‑checked news items labeled 'real' or 'fake' by the Politifact fact‑checking website, along with their retweet cascades; GossipCop \cite{gossipcopkaggle}, a larger dataset of entertainment news with similar annotations.

To address this limitation, we propose a lightweight enhancement: for each node in the propagation graph, we compute two classical graph-theoretic metrics: degree centrality and local clustering coefficient, and concatenate them with the node's original BERT and profile embeddings. This explicit marking of the hub vs. community roles requires only a few lines of additional code, but increases macro F1 from 0.7753 to 0.8344 on the UPFD Politifact subset. Our approach not only improves detection accuracy, but also provides an interpretable and easily reproducible template for combining graph metrics in other information‑diffusion contexts.

\section{Research Questions}

To test this hypothesis, we propose a lightweight enhancement to the UPFD framework by concatenating degree centrality and local clustering coefficient to the node embeddings. We evaluate our approach on two datasets, Politifact and GossipCopand,  aim to answer the following research questions:

RQ1: Can the incorporation of explicit topological features (degree centrality and local clustering coefficient) improve fake news detection performance in GNN-based frameworks such as UPFD?

RQ2: Does the effectiveness of these structural features vary across datasets of different domains (e.g., political news vs. entertainment news)?

These questions aim to explore both the utility and generalizability of structural graph features in the context of misinformation detection.

\section{Dataset}

UPFD \cite{ upfd}utilizes two major datasets, Politifact \cite{Politifact} and GossipCop, both derived from the FakeNewsNet repository. Politifact consists of 314 fact-checked political news items labeled as real or fake alongside retweet cascades comprising around 41,000 nodes. GossipCop \cite{gossipcopkaggle}, larger in scale, contains 5,464 entertainment news articles with over 314,000 nodes representing user interactions. These datasets enable comprehensive analysis of news credibility, integrating both textual and propagation-based social contexts.

\section{Methodology}

To address the limitations of the original UPFD framework, particularly its implicit treatment of graph topology, we propose a strengthened model that explicitly incorporates node-level structural signals and leverages a more expressive graph encoder architecture.

\subsubsection{Topological Feature Augmentation}

In standard UPFD, node features are composed of user profile embeddings and textual representations (e.g., BERT-encoded tweet history). While effective, this approach omits explicit indicators of a node’s role in the propagation graph. Motivated by classical social network analysis, we enhance each node representation by appending two interpretable topological features:

\textbf{Degree Centrality:}  A normalized scalar indicating the importance of a user as a hub within the graph.

\textbf{Local Clustering Coefficient:} A measure of local cohesiveness, reflecting how tightly the user is embedded within a community.

These metrics are computed using the NetworkX library after converting the PyTorch Geometric graph to a NetworkX graph object. The resulting values are concatenated to the original node feature vectors prior to model training. This design philosophy aligns with flexible tuning paradigms that unbind the tuner from the backbone \cite{jiang2023restuning}. Furthermore, it resonates with recent approaches in in-context meta-optimized fusion for multi-task adaptation \cite{shao2026icmfusion}.

The efficacy of utilizing explicit features for predictive modeling has been validated across diverse complex systems beyond computer science. For example, Li et al. successfully applied predictive methods to estimate the peak deviatoric stress of gravels \cite{li2024gravel}. Building on this, they further demonstrated the capability to predict complex stress-strain behaviors in granular materials \cite{li2025gravel}. Drawing inspiration from these cross-disciplinary successes in modeling physical stress, we apply explicit topological stress indicators (centrality) to model the "pressure" points in information diffusion networks.

\subsubsection{Graph Neural Architecture}

To exploit the enriched feature space, we employ a more expressive graph classification architecture based on the Graph Isomorphism Network (GIN) and an attention-based global readout:

\textbf{Encoder:} A single-layer GINConv module is used to transform the node features. Compared to GCN or GraphSAGE, GIN has been shown to have stronger discriminative power, particularly in distinguishing graph structures.

\textbf{Readout:} Instead of using mean or sum pooling, we utilize Global Attention Pooling (AttentionalAggregation), which computes a learned importance score for each node and aggregates them accordingly. This allows the model to emphasize structurally or semantically critical nodes during the graph-level classification.

\textbf{Classifier Head:} The pooled graph embedding is passed through a fully connected feedforward layer, which includes batch normalization and dropout for regularization, followed by a softmax activation for binary classification.

Formally, let $\mathbf{X} \in \mathbb{R}^{n \times d}$ be the matrix of node features, including topological enhancements. The model computes:

\[
\mathbf{H}^{(1)} = \text{ReLU}(\text{GINConv}(\mathbf{X}, \mathbf{A}))
\]
\[
\mathbf{z} = \text{AttnPool}(\mathbf{H}^{(1)}) = \sum_{i=1}^n \alpha_i \mathbf{h}_i
\]
\[
\hat{y} = \text{Softmax}(f(\mathbf{z}))
\]

where $\alpha_i$ are attention weights computed via a learnable scoring function over node embeddings.

\subsection{Training Configuration}

The model is trained using the Adam optimizer with a learning rate of $e^{-3}$ and weight decay of $5e^{-4}$. We use a batch size of $64$ and train for $50$ epochs. Input node features are initialized using pretrained BERT embeddings, with topological features appended during the preprocessing step.

This architecture is implemented using the PyTorch Geometric framework and evaluated on the Politifact and GossipCop datasets. Empirical results demonstrate consistent improvements over UPFD baselines using GCN and GraphSAGE encoders, validating the effectiveness of our topological feature integration and encoder design.

\subsubsection{Feature Importance Analysis}

To better understand the relative contribution of node features and graph structure to model performance, we implement an ablation-based \textit{feature importance analyzer}. This module evaluates model accuracy under three settings:

\textbf{Original Model:} Trained and tested on real node features and graph structure.
     
\textbf{Feature-Only Model:} Retains node features but replaces the graph topology with randomized edges.
 
\textbf{Structure-Only Model:} Retains the original graph topology but replaces node features with Gaussian noise.

By measuring the accuracy degradation under each ablation, we quantify the dependency of the model on content versus structure. Results on the Politifact and GossipCop datasets indicate that the importance of node features and graph structure varies across domains. For instance, in politically-oriented content (Politifact), user-specific textual signals tend to dominate, whereas in entertainment-related propagation (GossipCop), structural diffusion patterns may provide additional discriminative cues.

\subsection{Baseline Models}

To provide a fair and interpretable evaluation, we compare our enhanced model against three commonly used GNN-based baselines: \textbf{GCN}, \textbf{GraphSAGE}, and \textbf{GAT}. Each model adopts a standard architecture composed of a single graph convolution layer followed by a global max pooling operation and a linear classification head.

These baselines differ from our proposed method in several key aspects:

 \textbf{No topological features:} Node representations are learned solely from profile or text-based features, without explicit use of structural metrics such as degree or clustering coefficient.
 
\textbf{Simpler aggregation:} All models use global max pooling for graph-level representation, whereas our method applies attention-based readout to weight node contributions adaptively.

\textbf{Weaker encoder capacity:} Compared to our GIN-based encoder, these models are known to have limited expressive power in distinguishing graph structures.

In some configurations, the baseline models optionally concatenate the news post embedding with the pooled graph embedding. However, the models still rely on implicit structure learning, lacking the interpretable and lightweight enhancements introduced in our approach.

\section{Analysis}

\subsubsection{Performance}

\begin{figure}[htbp]
    \centering
    \includegraphics[width=0.5\textwidth]{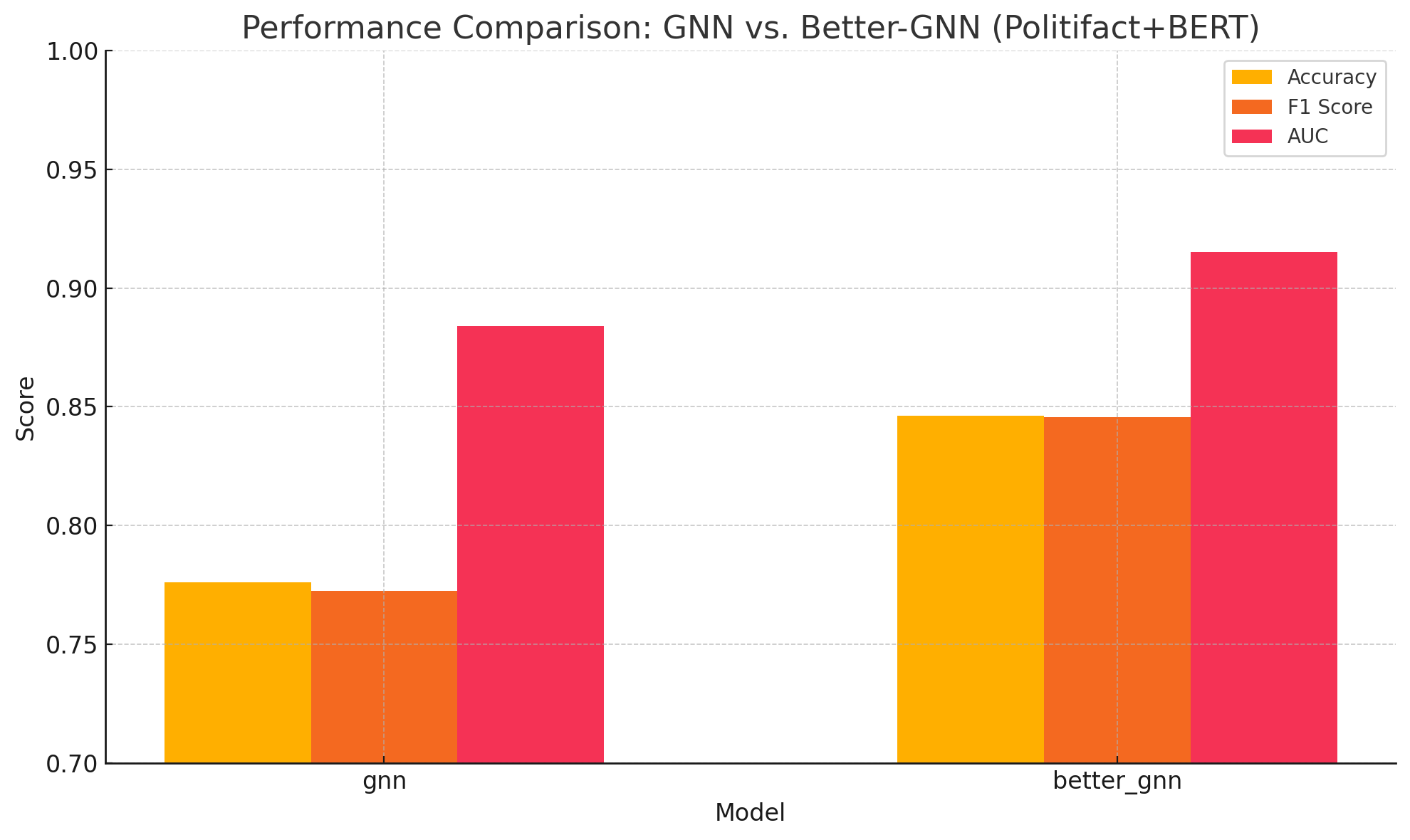}
    \caption{Performance comparison between GNN and Better-GNN on the Politifact dataset with BERT features.}
    \label{fig:politifact_performance}
\end{figure}

\begin{figure}[htbp]
    \centering
    \includegraphics[width=0.5\textwidth]{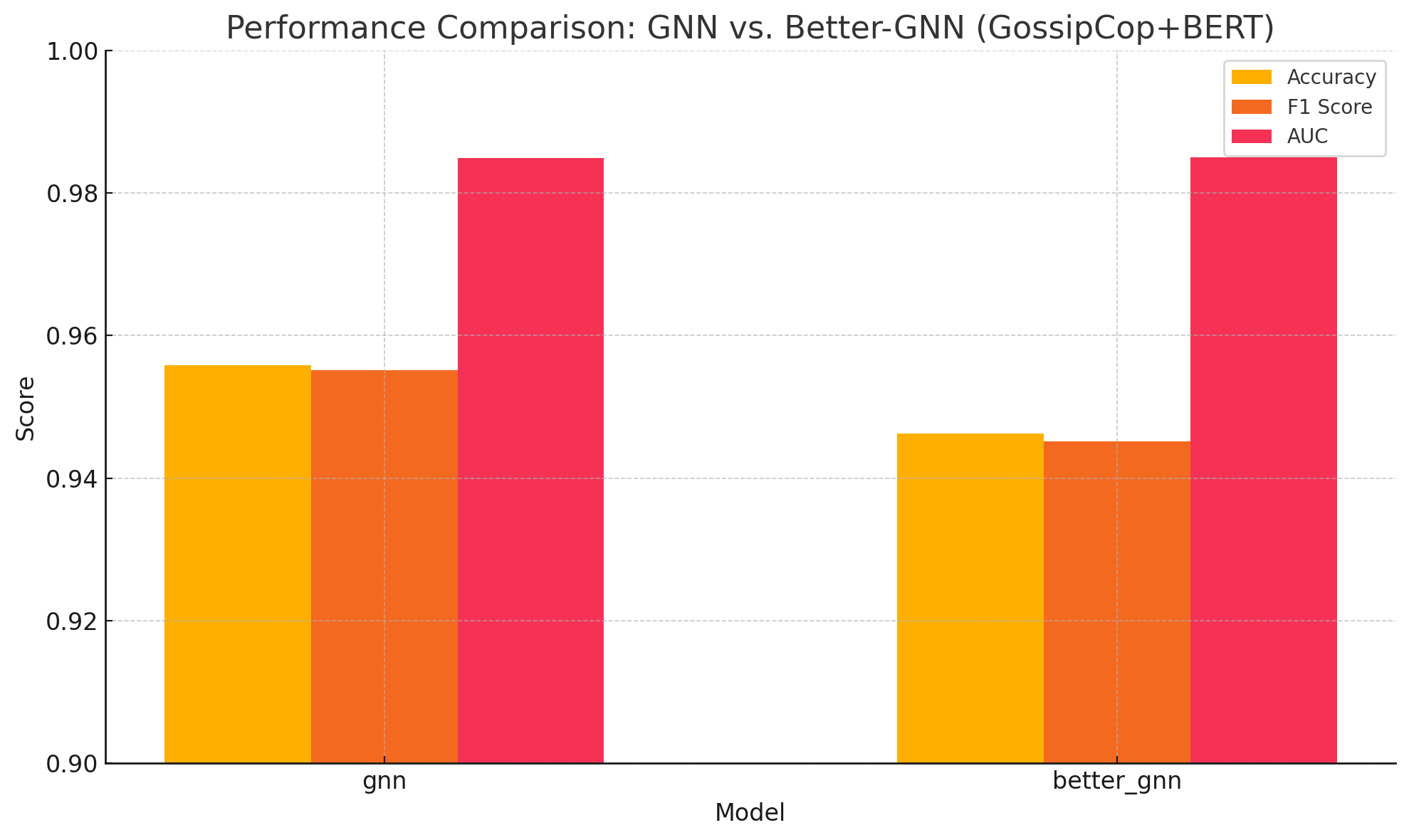}
    \caption{Performance comparison between GNN and Better-GNN on the GossipCop dataset with BERT features.}
    \label{fig:gossipcop_performance}
\end{figure}

\begin{figure*}[htbp]
    \centering
    \includegraphics[width=1\textwidth]{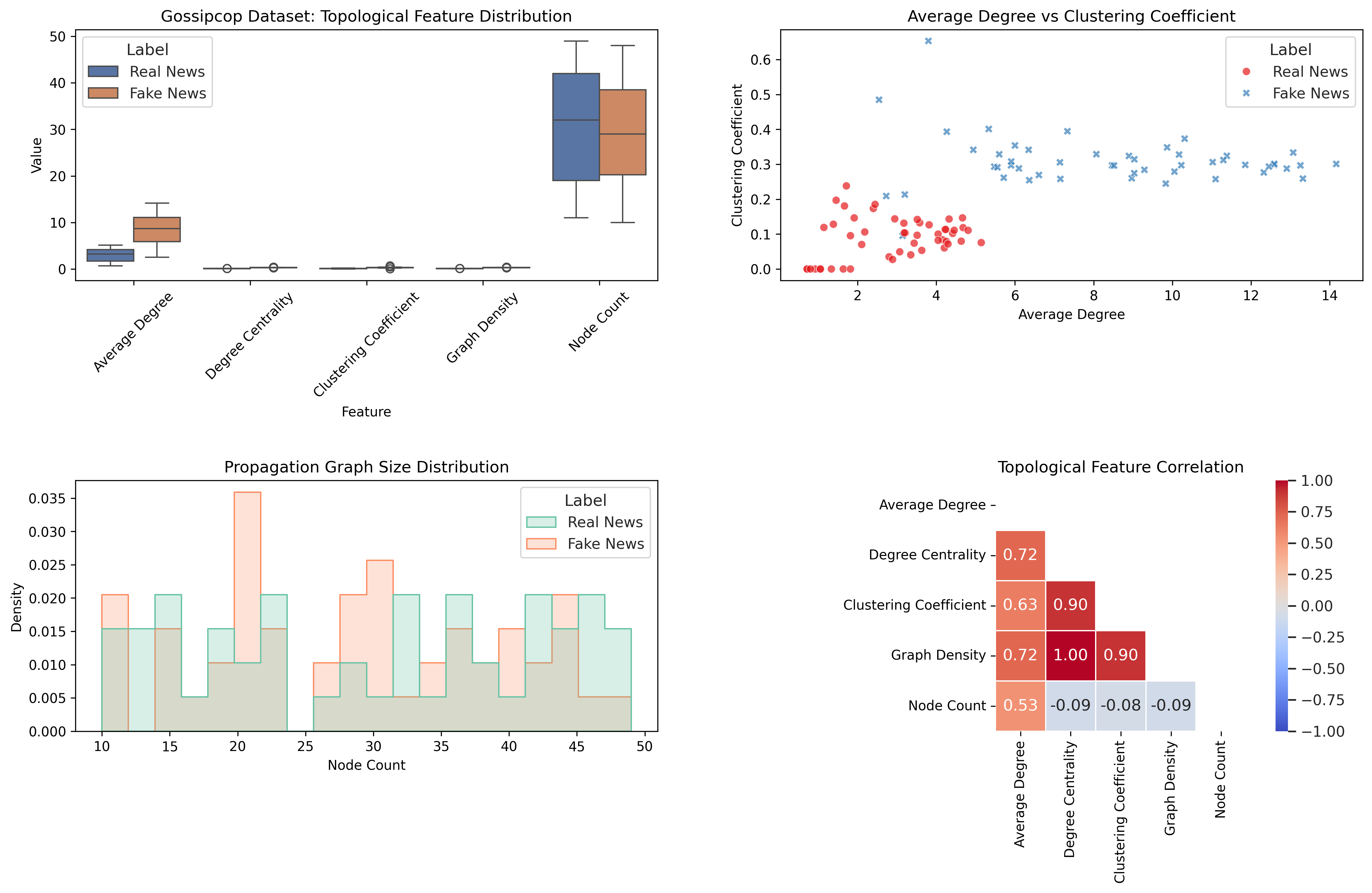}
    \caption{GossipCop dataset Topology Feature Analysis.}
    \label{fig:gossipcop_topo}
\end{figure*}

\begin{figure*}[htbp]
    \centering
    \includegraphics[width=1\textwidth]{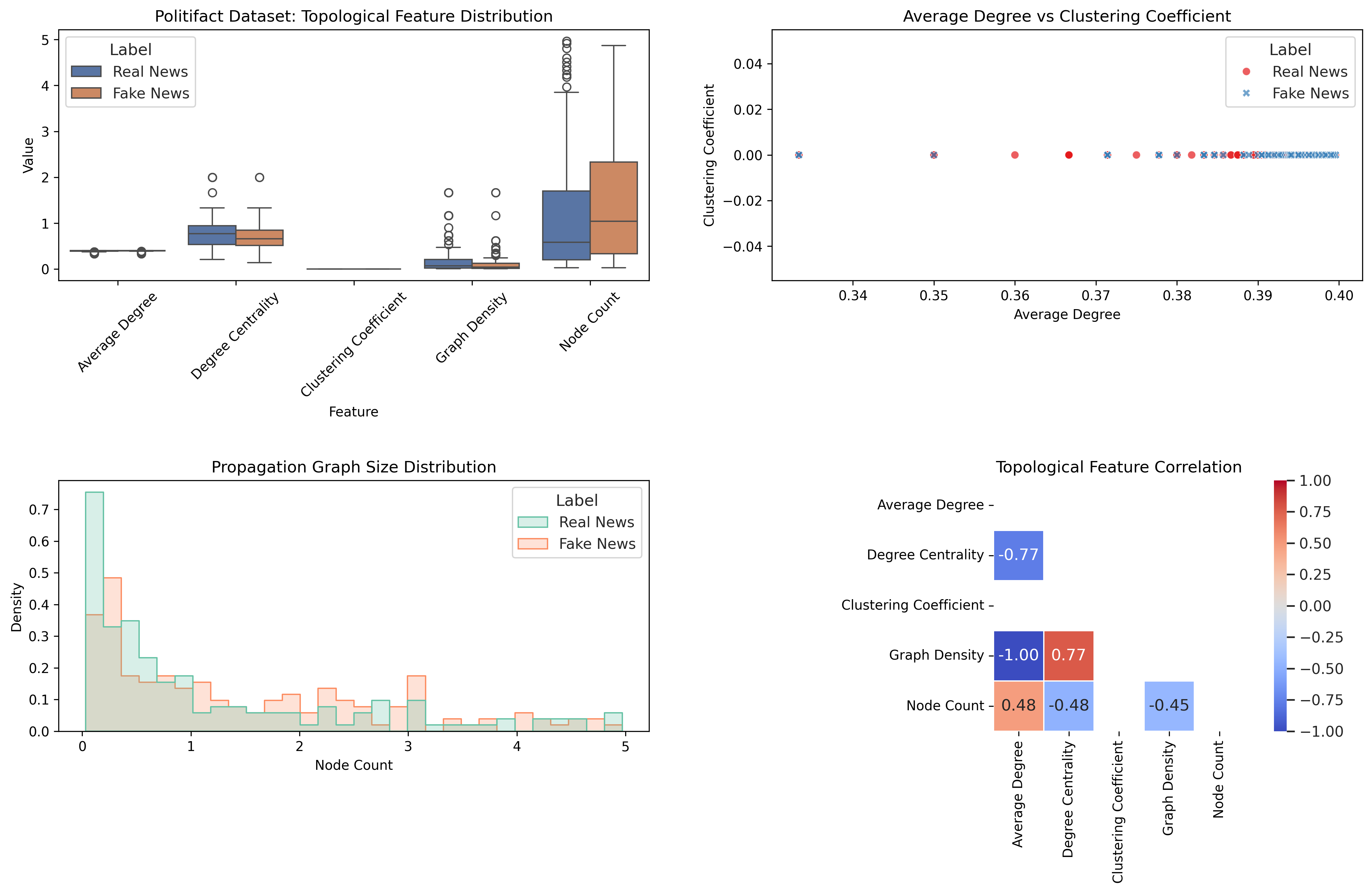}
    \caption{Politifact dataset Topology Feature Analysis.}
    \label{fig:gossipcop_topo}
\end{figure*}

To better understand the impact of our proposed topological enhancements, we visualize the performance of the baseline GNN model and our improved Better-GNN on two datasets: Politifact and GossipCop, both using BERT as the node feature encoder.

Figure~\ref{fig:politifact_performance} shows the evaluation results on the Politifact dataset. The Better-GNN model achieves consistent improvements over the standard GNN across all metrics. Specifically, macro F1 improves from 0.7725 to 0.8455, and AUC increases from 0.8839 to 0.9152. This indicates that incorporating explicit topological features such as degree centrality and local clustering coefficient significantly enhances the model’s ability to distinguish between fake and real political news, where propagation structure carries strong discriminative cues.

In contrast, Figure~\ref{fig:gossipcop_performance} illustrates the results on the GossipCop dataset. Here, the performance gains from Better-GNN are marginal or even slightly lower in terms of accuracy and F1 score. While AUC remains nearly identical (0.9850), the baseline GNN achieves slightly better F1 (0.9551 vs. 0.9451). This suggests that in entertainment-related content, where propagation structure may be less informative or more homogeneous, the added structural features provide less benefit and may even introduce noise or redundancy.

Overall, these results highlight the importance of dataset characteristics in determining the effectiveness of graph-based enhancements. While Better-GNN is particularly advantageous in domains where structural variation reflects truthfulness (e.g., political discourse), it may offer limited advantages in domains dominated by content-based signals.

The observed performance difference between the two datasets can be explained by domain-specific propagation behaviors. In political news (e.g., Politifact), misinformation tends to spread within ideologically homogeneous communities, resulting in densely connected local structures with high clustering coefficients and prominent hub users. These patterns are strongly correlated with the truthfulness of the news content and are effectively captured by structural features like degree centrality and local clustering.

Moreover, political misinformation often employs vague or ambiguous language, making it harder to detect through content-based features alone. This increases the reliance on propagation-based signals to enhance model discrimination. By contrast, in entertainment news (e.g., GossipCop), content tends to be more sensational and easier to detect using textual features, while the propagation patterns are flatter, noisier, and less community-driven. As a result, structural signals provide less marginal benefit in such cases and may introduce redundancy.

This analysis suggests that structural enhancements are particularly beneficial in domains where misinformation spreads through cohesive and ideologically polarized networks, reinforcing the case for domain-aware design in graph-based misinformation detection systems.

\subsubsection{Topology Feature Analysis}

We compare the structural properties of news propagation graphs between the Politifact and GossipCop datasets using five key topological features: average degree, degree centrality, clustering coefficient, graph density, and node count. These features capture both local connectivity and global graph structure.

From the boxplots (top-left), we observe the following: In Politifact, fake news graphs tend to have lower degree centrality and lower graph density, while their node count (i.e., the size of the propagation network) is generally higher than that of real news. This suggests that political misinformation may spread through larger but sparser social networks. In GossipCop, fake news exhibits a significantly higher average degree and clustering coefficient than real news, indicating denser and more tightly knit local propagation structures in entertainment-related misinformation. These trends suggest that political fake news tends to spread wide but thin, whereas entertainment fake news spreads more compactly within local communities.

The scatter plots (top-right) visualize the relationship between average degree and clustering coefficient: In Politifact, most graphs cluster around low values on both axes, with limited separation between real and fake news. This indicates weak structural signals in terms of local density. In GossipCop, fake news graphs are clearly separated from real news, occupying higher ranges on both axes. This shows that topological metrics can better distinguish fake news in the entertainment domain.

The node count distribution (bottom-left) further highlights the differences: In Politifact, fake news graphs are noticeably larger in size than real news, showing a long tail in the number of participating users. In GossipCop, both real and fake news exhibit comparable sizes, though fake news slightly dominates the mid-size range.

The correlation heatmaps (bottom-right) reveal internal redundancy: In both datasets, average degree, graph density, and clustering coefficient are strongly correlated ( $\rho$ > 0.6), suggesting that they capture similar structural patterns. Node count, however, shows low correlation with other features, indicating it measures a distinct aspect of the graph (scale, not density). This suggests that a subset of features (e.g., node count + degree centrality) may be sufficient to capture both scale and density information for classification.

GossipCop shows stronger and more separable topological signals for fake news detection, especially in compactness-related metrics. Politifact highlights the importance of scale (node count), rather than dense connectivity. These insights support feature-aware and domain-aware design of GNN models, such as selectively enhancing structure-aware modules based on dataset type.

\section{Related Work}

Despite its strong performance, UPFD leaves all graph‑level structural learning to the GNN, without exposing any explicit topological signals, such as which users function as high-degree hubs or which communities are tightly knit. This necessity of integrating explicit statistical indicators mirrors methodological discussions in broader analytical fields, where relying solely on implicit data often fails to capture the full scope of competitive or structural dynamics~\cite{HeJingwen1, HeJingwen2, HeJingwen3, HeJingwen4}.

Recent studies have shown that large language models are sensitive not only to prompt semantics, but also to seemingly superficial linguistic cues. For example, Cai et al.~\cite{cai2025does} systematically evaluate the impact of prompt politeness and tone on the outputs of modern LLMs, including GPT-, Gemini-, and LLaMA-family models, and observe non-trivial variations in answer quality and consistency. These findings suggest that LLM behavior can be influenced by non-semantic prompt factors, raising concerns about robustness and reproducibility.

\section{Conclusion}

In this study, we introduced a lightweight yet effective enhancement to existing GNN-based fake news detection frameworks by explicitly incorporating node-level topological features, namely degree centrality and clustering coefficient. Through extensive experiments on two benchmark datasets—Politifact and GossipCop—we demonstrated that these structural features not only improve performance in politically-oriented misinformation detection but also offer a reproducible and interpretable modeling strategy.

Furthermore, our comparative analysis reveals that the utility of topological features varies by domain: while structural metrics significantly boost performance in political contexts characterized by polarized communities and sparse propagation, their benefit is less pronounced in entertainment domains where content-based cues are dominant.

These findings emphasize the need for domain-aware and feature-aware GNN design in misinformation detection tasks. Future work may explore adaptive feature selection based on dataset structure, or extend this approach to multimodal fake news settings involving images and videos.

Future work may explore adaptive feature selection based on dataset structure, or extend this approach to multimodal fake news settings. Specifically, training-free approaches for event-aware video anomaly detection could be adapted to identify manipulated news footage \cite{shao2025eventvad}. To handle the scale of multimedia data, in-network tensor fusion techniques offer a path for scalable feature indexing \cite{wang2025vifusion}. Finally, learning robust matching with selective mixture-of-experts could provide critical cues for verifying visual consistency in the wild \cite{wang2025learning}.

\bibliography{anthology,custom}
\bibliographystyle{acl_natbib}

\appendix



\end{document}